# Shaping Integrity: Why Generative Artificial Intelligence Does Not Have to Undermine Education

Myles Joshua Toledo Tan[1,2,3,4,5,6,7,*], Nicholle Mae Amor Tan Maravilla[7]

[1] Department of Electrical and Computer Engineering, Herbert Wertheim College of Engineering, University of Florida, Gainesville, Florida, 32611, United States of America

[2] Department of Epidemiology, College of Public Health & Health Professions and College of Medicine, University of Florida, Gainesville, Florida, 32610, United States of America

[3] Biology Program, College of Arts and Sciences, University of St. La Salle, Bacolod City, Negros Occidental, 6100, Philippines

[4] Department of Natural Sciences, College of Arts and Sciences, University of St. La Salle, Bacolod City, Negros Occidental, 6100, Philippines

[5] Department of Chemical Engineering, College of Engineering and Technology, University of St. La Salle, Bacolod City, Negros Occidental, 6100, Philippines

[6] Department of Electronics Engineering, College of Engineering and Technology, University of St. La Salle, Bacolod City, Negros Occidental, 6100, Philippines

[7] Yo-Vivo Corporation, Bacolod City, Negros Occidental, 6100, Philippines

\* Correspondence
Myles Joshua Toledo Tan, mylesjoshua.tan@medicine.ufl.edu

## Introduction

The integration of generative artificial intelligence (GAI) in education has been met with both excitement and concern. According to a 2023 survey by the World Economic Forum, over 60% of educators in advanced economies are now using some form of artificial intelligence (AI) in their classrooms, a significant increase from just 20% five years ago (World Economic Forum, 2023). The rapid adoption of AI technologies in education highlights their potential to revolutionize the learning experience. AI tools, such as intelligent tutoring systems and adaptive learning platforms, offer personalized educational experiences that can meet the unique needs of each student. However, with this potential comes significant ethical concerns, particularly regarding academic integrity.

The International Center for Academic Integrity reported that 58% of students admitted to using AI tools to complete assignments dishonestly, highlighting the urgency of addressing these ethical concerns (International Center for Academic Integrity, 2023). This statistic underscores a critical issue: while AI has the potential to enhance education, its misuse can undermine the very foundations of academic integrity. The rise of AI technology has raised concerns about academic integrity. With tools that can generate text, solve problems, and even assist with research, students may find it easier to engage in plagiarism or other forms of cheating. This shift challenges traditional educational values, as it blurs the lines between original work and AI-generated content (Mohammadkarimi, 2023). Curriculum designers are thus faced with the challenge of integrating AI in ways that uphold ethical standards and promote genuine learning. This requires balancing the innovative potential of AI tools with a commitment to academic integrity, ensuring that technology enhances rather than undermines the educational experience.

To navigate this landscape responsibly, it is essential to revisit established ethical frameworks and educational theories. The ethical principles guiding our use of technology in education have remained consistent, even as the tools themselves have evolved. By referencing seminal works and foundational theories, we can demonstrate that the core values of honesty, fairness, and responsibility are timeless. For example, deontological ethics, as articulated by Immanuel Kant, emphasizes the importance of adhering to moral principles such as honesty and integrity, rather than the consequences of actions (Kant, 1785). In the context of AI in education, deontological ethics would require that the use of AI respects fundamental moral principles. For example, it would be crucial to ensure that AI systems are designed and implemented in ways that uphold students' rights to privacy, ensure fairness, and avoid deception. Adhering to these principles would be seen as morally obligatory, regardless of the potential benefits or drawbacks of AI in educational settings. Similarly, consequentialism, as articulated by John Stuart Mill, evaluates actions based on their outcomes. Mill's version of consequentialism, known as utilitarianism, argues that the best actions are those that promote happiness or better well-being. In the context of AI in education, applying Mill's consequentialist principles would involve assessing how the use of AI impacts educational outcomes. If AI can be used to enhance learning, provide personalized educational experiences, or address inequalities and inequities in education, then its use would be considered morally justified according to Mill's framework, as it promotes overall well-being and positive outcomes for students.

These ethical frameworks provide a robust foundation for the responsible use of GAI in modern educational settings. Moreover, educational theories such as constructivist learning and Self-Determination Theory (SDT) offer valuable insights into how AI can be used to enhance learning. Constructivist learning theory posits that students construct knowledge through active engagement with content, a process that can be greatly facilitated by AI tools. This approach emphasizes the importance of students' engagement in hands-on activities and interactions, which help them construct meaningful connections with new information (Hein, 1991). AI tools can significantly enhance this constructivist approach by providing personalized and interactive learning experiences. SDT, on the other hand, emphasizes the importance of autonomy, competence, and relatedness in fostering intrinsic motivation among students (Deci & Ryan, 2000). Integrating AI tools that align with the principles of SDT can help create a more engaging and supportive learning environment among students.

This discussion will explore how GAI can be integrated into education in ways that support rather than erode academic integrity. By examining the ethical frameworks of deontological ethics and consequentialism, and educational theories like constructivist learning and SDT, we will argue that AI, when used responsibly, can enhance digital literacy, foster intrinsic motivation, and support genuine knowledge construction. The principles discussed in older foundational papers remain relevant, proving that ethical guidelines established decades ago still hold value in today's technologically advanced classrooms (Floridi & Taddeo, 2016; Ryan & Deci, 2017).

The goal is to illustrate that the ethical use of GAI in education not only preserves but can also enhance academic integrity. Through responsible integration and ethical education, AI can empower students to become motivated, ethical, and engaged learners, well-prepared for the complexities of the modern world. By grounding our arguments in established ethical and educational theories, we can provide a comprehensive framework for understanding the potential benefits and challenges of AI in education.

**Navigating the Disruptive Impact of Generative Artificial Intelligence on Assessment**

The integration of GAI in education raises significant concerns about its potential to disrupt traditional assessment methods. The ability of GAI to generate essays, problem solutions, and even creative works has sparked fears of plagiarism and academic dishonesty, challenging conventional forms of evaluation such as take-home exams, essays, or homework assignments. These concerns are valid, as the ease with which students can use AI-generated content without truly engaging in the learning process threatens to undermine academic integrity (Popenici and Kerr, 2017).

However, the disruptive nature of GAI also presents an opportunity to reimagine assessment practices in ways that prioritize authentic learning and deeper understanding. The rise of AI necessitates a shift away from traditional assessments focused on rote memorization and information recall, toward more authentic assessment methods that require students to demonstrate higher-order thinking skills. For example, project-based tasks, real-world problem-solving activities, oral presentations, and open-ended assignments that demand personal reflection and original insights can reduce the likelihood of misuse and encourage students to engage meaningfully with course material (Borenstein and Howard, 2020).

Furthermore, GAI can play a constructive role in formative assessment by providing personalized feedback throughout the learning process. AI-driven tools can help students revise drafts, practice skills, and receive immediate guidance on areas needing improvement, fostering a deeper connection to the material. This approach transforms GAI from a potential threat to a valuable asset that supports continuous learning and skill development. Additionally, incorporating self-assessment and metacognitive practices, where students reflect on their progress and learning strategies, can ensure that AI augments rather than diminishes students' active participation in their education.

It is also essential to address the ethical considerations involved in using AI for assessment. Concerns such as data privacy, algorithmic bias, and the fairness of AI-generated evaluations must be taken seriously (Borenstein and Howard, 2020). Developing clear institutional policies that set boundaries on acceptable AI use in assessments can help maintain fairness and transparency. These policies should include guidelines for combining AI insights with human judgment to ensure that assessments reflect not only the outputs o AI but also the educator's understanding of the student's abilities and efforts.

By embracing these strategies, educators and institutions can harness the potential of GAI to enhance assessments while maintaining academic integrity. This balanced approach allows for the responsible integration of AI in education, ensuring that it supports meaningful learning experiences and prepares students to navigate an AI-driven world with integrity.

**Constructivist Learning Theory: Enhancing Knowledge Construction**

Constructivist learning theory posits that learners construct knowledge through experiences and reflections, actively engaging with content to build understanding. GAI, with its advanced capabilities, aligns well with this theory, offering tools that promote exploration, interaction, and personalized learning paths. Contrary to the belief that AI erodes academic integrity, some scholars argue that AI, when used thoughtfully, has the potential to enhance educational experiences by providing personalized learning opportunities and supporting students' individual learning needs (Weller, 2020). While Weller does not claim that AI inherently fosters critical thinking or deeper understanding, his discussion highlights the potential of AI in educational settings, suggesting that it could complement traditional teaching methods to improve learning outcomes.

GAI tools, such as intelligent tutoring systems and adaptive learning platforms, provide students with tailored educational experiences. These systems analyze individual learning patterns and adapt content to meet specific needs, ensuring that students engage with material at an appropriate level of difficulty (Woolf, 2010). For instance, an AI-powered math tutor can identify a student's weaknesses in algebra and offer targeted exercises to address these gaps. This personalized approach not only supports knowledge construction but also encourages students to take ownership of their learning journey (Shute & Zapata-Rivera, 2012).

In a classroom setting, imagine a high school history class studying the Industrial Revolution. The educator integrates a GAI tool that generates interactive timelines and simulations based on historical data. Students can manipulate variables within these simulations to observe the effects on industrial growth, labor conditions, and economic development. Through this exploration, they construct a deeper understanding of the era's complexities. Instead of passively receiving information, students actively engage with content, reflecting on the consequences of different actions and decisions (Kumar et. al., 2024).

Another example is in language arts, where a GAI tool assists students in creative writing. By analyzing a student's writing style and providing real-time feedback on grammar, tone, and narrative structure, the AI helps students refine their skills (Song & Song, 2023). Additionally, it can suggest plot developments or character traits, sparking students' creativity and encouraging them to think critically about their stories. This interactive process supports constructivist principles by allowing students to experiment, reflect, and build upon their ideas (Bereiter & Scardamalia, 1989).

Critics argue that AI tools may encourage academic dishonesty by making it easier for students to produce work with minimal effort. However, this perspective overlooks the potential for AI to promote genuine learning when used appropriately. Rather than replacing student effort, AI can enhance the learning process by offering personalized support, immediate feedback, and adaptive content, which fosters deeper engagement and learning outcomes (Nazaretsky et al., 2022). For instance, in a science class, AI-powered lab assistants can guide students through virtual experiments, providing explanations and prompting them to hypothesize, analyze data, and draw conclusions. Such interactions encourage active learning and promote a deeper understanding of scientific concepts and processes, rather than merely supplying answers (de Jong & van Joolingen, 1998). Additionally, as Al Daraysh (2023) notes, AI tools designed with input from educators help align the technology with pedagogical objectives, embedding ethical considerations to reduce the risk of academic dishonesty. Furthermore, it is important to acknowledge that AI is transforming science education and pedagogy, and the ethical implementation of these tools must reflect this shift to support genuine learning experiences while safeguarding academic integrity (Holstein et al., 2018; Erduran, 2023).

Moreover, GAI can facilitate collaborative learning, another key aspect of constructivist theory. In a project-based learning environment, students can use AI tools to collaboratively develop presentations or reports. AI can assist by organizing information, suggesting relevant sources, and providing feedback on the clarity and coherence of their work (Kreijns et al., 2003). This collaborative process encourages students to engage in dialogue, share perspectives, and build knowledge collectively.

To further illustrate, consider a classroom where students are tasked with developing a business plan. An AI tool can generate market analysis reports, financial projections, and strategic recommendations based on input from the students. As they interact with the AI and with each other, they learn to critically evaluate information, make informed decisions, and adapt their plans. This dynamic, interactive process is at the heart of constructivist learning, fostering not only knowledge construction but also critical thinking and problem-solving skills (Jonassen, 1995).

At present, there are multiple AI powered tools that are being used by most students that have significant potential to enhance a constructivist learning experience. One example is the ChatGPT. According to Rasul et.al (2023), ChatGPT supports the constructivist principle that learners construct their own understanding of knowledge by enabling students to explore and experiment with ideas, ask questions, and receive immediate feedback. This interactive engagement helps students to deeply connect with the content, refine their comprehension, and apply their learning in meaningful ways, ultimately enriching their educational experience.

Also, according to Mota-Valtierra et al (2019), a constructivist approach is a great fit for teaching AI topics because it emphasizes building on prior knowledge and encouraging active learning. Their article outlines an innovative approach to teaching artificial intelligence (AI) through a constructivist methodology, specifically focusing on multilayer perceptrons (MLPs). After implementing it in different majors, the statistical analysis underscores the success of the proposed course methodology in enhancing student learning and providing a more consistent educational experience. The increase in average grades and the reduction in standard deviation highlight the effectiveness of the approach in improving both individual performance and overall learning outcomes.

In conclusion, GAI aligns with constructivist learning theory by providing tools that facilitate exploration, interaction, and personalized learning. Rather than promoting dishonesty, AI can enhance academic integrity by supporting genuine learning experiences. Through personalized feedback, interactive simulations, and collaborative projects, AI empowers students to take an active role in their education, constructing knowledge in meaningful and engaging ways. By embracing these technologies, educators can create enriching learning environments that prepare students for the complexities of the modern world (Papert & Harel, 1991).

**The Ethics of Artificial Intelligence: Responsible Use and Digital Literacy**

The rise of GAI in education has sparked discussions on its ethical implications and the importance of fostering digital literacy. By examining ethical frameworks such as deontological ethics and consequentialism, we can argue that responsible use of GAI in the classroom can enhance students' digital literacy and prepare them to navigate the digital world ethically and effectively (Floridi & Taddeo, 2016; Stahl, 2012).

Deontological ethics, which focuses on adherence to moral rules or duties, provides a foundation for integrating AI responsibly in education. This framework emphasizes the importance of principles such as honesty, fairness, and respect for others (Kant, 1785). In the context of GAI, this means ensuring that AI tools are used to support and enhance learning rather than replacing students' efforts or promoting dishonesty.

For instance, in a high school history class studying the Industrial Revolution, an AI tool can generate interactive timelines and simulations based on historical data. Educators can emphasize the importance of using these tools ethically, encouraging students to engage with the material thoughtfully and critically. By adhering to principles of honesty and integrity, students learn to use AI as a supplementary resource that enhances their understanding rather than as a shortcut to completing assignments (Johnson, 2020).

Consequentialism, as articulated by John Stuart Mill in *Utilitarianism* (1861), evaluates the morality of actions based on their outcomes. While Mill did not discuss AI, the principles of this framework can still be applied to contemporary debates about its use in education. By aiming to maximize positive outcomes—such as enhanced learning, critical thinking, and digital literacy—educators and curriculum designers can advocate for the responsible integration of AI. Emphasizing these benefits underscores how AI tools can contribute to better educational results and foster more informed digital citizens.

In a language arts classroom, for example, a GAI tool can assist students in creative writing by providing real-time feedback on grammar, tone, and narrative structure. Educators can guide students to use this feedback to improve their writing skills, fostering a deeper understanding of language and storytelling. The positive outcomes of enhanced writing abilities and critical engagement with AI tools illustrate the ethical benefits of responsible AI use (Borenstein & Howard, 2020).

To further promote digital literacy, it is crucial to educate students and educators on the ethical use of AI tools. This involves teaching them to understand how AI works, the potential biases and limitations of AI systems, and the importance of using AI responsibly (Brey, 2012). By fostering a culture of digital literacy, educators empower students to navigate the digital world with a critical and ethical mindset.

Consider a science class where an AI-powered lab assistant guides students through virtual experiments. Educators can use this opportunity to discuss the ethical considerations of AI in scientific research, such as data privacy, bias, and the importance of accurate data interpretation. By engaging in these discussions, students develop a nuanced understanding of the role of AI in science and the ethical responsibilities of using AI in research (Floridi, 2013).

Moreover, collaborative projects can further enhance digital literacy and ethical awareness. In a project-based learning environment, students can use AI tools to develop presentations or reports collaboratively. Educators can emphasize the importance of ethical collaboration, such as giving credit to sources, avoiding plagiarism, and ensuring that all team members contribute fairly. This approach not only enhances students' digital literacy but also instills ethical values that are essential in the digital age (Ess, 2015).

For instance, in a business class where students are tasked with developing a business plan, an AI tool can generate market analysis reports and financial projections. Educators can guide students to critically evaluate the AI-generated data, discuss the ethical implications of using AI in business decision-making, and ensure transparency and accountability in their work. This process helps students understand the ethical dimensions of AI and develop skills to use AI responsibly in their future careers (Mittelstadt et al., 2016).

The ethical frameworks of deontological ethics and consequentialism provide valuable insights into the responsible use of GAI in education. By emphasizing the importance of principles such as honesty, fairness, and positive outcomes, educators can foster digital literacy and ethical awareness among students. Teaching students to understand and navigate the ethical implications of AI tools prepares them to contribute positively to the digital world, ensuring that they use AI to enhance learning and uphold ethical standards. Through responsible AI integration and ethical education, we can create a generation of digitally literate and ethically aware individuals ready to thrive in a technologically advanced society (Moor, 1985).

The integration of AI in education holds great promise for enhancing learning experiences but raises profound ethical questions. The need for careful ethical reflection is underscored in *The Ethics of Artificial Intelligence in Education: Practices, Challenges, and Debates*, which argues that educators, researchers, and stakeholders must engage in ongoing dialogue to navigate the complexities of AI in educational contexts (Holmes and Porayska-Pomsta, 2022). Smuha (2022) points out that for AI in education to be ethically responsible, it must adhere to key principles such as fairness, accountability, and transparency. These principles are vital in mitigating biases and preventing AI from perpetuating or amplifying existing educational inequalities. Furthermore, the concept of Trustworthy AI, as discussed by Smuha, is crucial in ensuring that AI systems foster inclusivity and do not marginalize vulnerable student populations (Smuha, 2022). Similarly, Brossi et al. (2022) raise concerns about the uncertain impact of AI on learners' cognitive development and the risk of disempowering educators through over-automation of pedagogical processes, pointing to the need for ethical frameworks that avoid automating ineffective or inequitable practices.

Williamson (2024) expands on this by highlighting the socio-political context of AI in education, warning against the assumption that technological innovations are inherently beneficial. Instead, he emphasizes that AI must be viewed as a socially embedded tool that could exacerbate educational inequities if not critically examined. The potential for AI to impact power dynamics, access, and social equity necessitates that educators and policymakers rigorously reflect on its broader implications, including how AI systems might reinforce or challenge existing educational structures.

Mouta et al. (2023) offer a practical step forward in addressing these concerns through their participatory futures approach, which is designed to help educators ethically integrate AI into their teaching environments. By using the Delphi method to gather diverse perspectives, their study presents hypothetical future scenarios that help educators and stakeholders reflect on the broader implications of AI in education. This approach ensures that the benefits of AI are balanced with ethical considerations related to privacy, bias, and the societal impacts of AI on education, promoting a thoughtful and inclusive implementation of AI technologies.

Further supporting this ethical stance, the European Commission's *Ethics Guidelines for Trustworthy AI* (2019) lays out seven key requirements for Trustworthy AI, including human agency, privacy, transparency, and fairness. These guidelines align closely with the need to ensure that AI systems in education promote fairness and inclusivity, rather than exacerbating inequities in educational access and outcomes. The guidelines also emphasize the importance of continuous monitoring and accountability to ensure AI systems remain aligned with these ethical principles. By stressing the importance of transparency, diversity, and non-discrimination, these guidelines reinforce the participatory frameworks put forth by Mouta et al. (2023), which advocate for an inclusive, ethical approach to AI integration in education.

Further reinforcing these ethical considerations, Floridi et al. (2018) in their "AI4People" framework emphasize the importance of a principled approach to AI that integrates ethical foundations like beneficence, non-maleficence, autonomy, justice, and explicability. These principles align with the need for AI in education to promote well-being and inclusivity while avoiding harm, respecting user autonomy, ensuring fair access to AI benefits, and fostering transparency. The framework also highlights that the potential risks of AI can include the erosion of human agency and privacy, making it essential for educational AI systems to be designed in ways that support rather than undermine student autonomy and self-determination. By embedding these principles into the development and deployment of AI, educators and policymakers can more effectively navigate the ethical challenges posed by AI in educational contexts, ultimately fostering a "Good AI Society" that supports human flourishing.

**Self-Determination Theory: Fostering Intrinsic Motivation**

SDT posits that individuals are most motivated when their needs for autonomy, competence, and relatedness are met. GAI, with its capability to provide personalized feedback and tailored learning resources, can significantly support SDT by fostering intrinsic motivation among students. By empowering students to take control of their learning, AI can enhance engagement and academic integrity (Deci & Ryan, 2000; Ryan & Deci, 2017).

*Autonomy*

GAI can enhance students' sense of autonomy by offering them more control over their learning process. In a high school history class studying the Industrial Revolution, an AI tool can create interactive timelines and simulations. Students can explore these tools at their own pace, choosing which aspects of the Industrial Revolution to delve into more deeply. This self-directed exploration encourages students to take ownership of their learning, fostering a sense of autonomy (Reeve, 2006).

For example, a student interested in labor conditions during the Industrial Revolution might use the AI tool to simulate different labor policies and observe their impacts. This personalized exploration helps students develop a deeper understanding of historical complexities, driven by their own curiosity and interests (Niemiec & Ryan, 2009).

*Competence*

GAI tools can also support the need for competence by providing personalized feedback that helps students improve their skills and knowledge. In a language arts classroom, an AI-driven writing assistant can analyze a student's work and provide targeted feedback on grammar, tone, and narrative structure. This real-time, individualized feedback helps students understand their strengths and areas for improvement, fostering a sense of competence (Black & Deci, 2000).

Imagine a student writing a short story. The AI tool can suggest improvements in plot development and character interactions, guiding the student to refine their narrative. As students see their writing improve through this iterative process, they gain confidence in their abilities, which enhances their intrinsic motivation to engage with the subject matter (Vansteenkiste et al., 2004).

*Relatedness*

GAI can also facilitate relatedness by enabling collaborative learning and providing opportunities for meaningful interactions. In a project-based learning environment, AI tools can help students work together on presentations or reports. For instance, in a science class, an AI-powered lab assistant can guide groups of students through virtual experiments, encouraging collaboration and discussion (Ryan & Powelson, 1991).

Consider a group of students using AI to simulate a chemical reaction. The AI provides each group member with specific tasks and prompts them to share their findings and discuss results. This collaborative process fosters a sense of relatedness, as students work together to achieve common goals and learn from each other (Jang et al., 2016).

*Promoting Academic Integrity*

By fostering intrinsic motivation through autonomy, competence, and relatedness, GAI can also promote academic integrity. When students are genuinely interested and engaged in their learning, they are less likely to resort to dishonest practices. Personalized learning experiences make education more relevant and enjoyable, reducing the temptation to cheat (Deci et al., 1991).

In history class, for example, students using AI to explore the Industrial Revolution are likely to develop a genuine interest in the subject. This intrinsic motivation drives them to produce original work and engage deeply with the material. Similarly, in the language arts class, students motivated by the desire to improve their writing skills are more likely to take pride in their work and avoid plagiarism (Vansteenkiste & Ryan, 2013).

*Real-World Application*

In a business class where students develop business plans using AI-generated market analysis reports and financial projections, educators can emphasize the importance of ethical decision-making and transparency. The AI tool provides personalized insights, allowing students to explore various business strategies and their consequences. This hands-on learning approach fosters intrinsic motivation by making the subject matter relevant and engaging (Ryan & Deci, 2000).

For instance, a student interested in starting a sustainable business can use AI to analyze the environmental impact of different business models. This personalized exploration helps the student develop a deeper understanding of sustainability in business, driven by their own interests and values (Deci & Ryan, 2008).

GAI, by supporting the principles of SDT, can foster intrinsic motivation among students. Through personalized feedback and tailored learning resources, AI empowers students to take control of their learning, enhancing their sense of autonomy, competence, and relatedness. This intrinsic motivation not only increases engagement but also promotes academic integrity. By integrating AI tools in educational settings, educators can create enriching learning environments that prepare students for the complexities of the modern world, ensuring that they are motivated, ethical, and engaged learners (Ryan & Deci, 2019).

**Discussion: Generative Artificial Intelligence as a Catalyst for Enhancing Academic Integrity**

The integration of GAI in education has sparked significant debate regarding its impact on academic integrity. Critics argue that AI tools facilitate dishonesty by providing easy shortcuts for students to complete assignments. However, a closer examination of established educational theories and ethical frameworks reveals a different perspective. When used responsibly, GAI can foster intrinsic motivation, enhance digital literacy, and support constructivist learning principles, thereby promoting academic integrity rather than eroding it.

The integration of GAI in various educational fields, including computer science, engineering, medical education, and communication, is revolutionizing teaching and learning. The integration of AI technologies in computer science education, particularly through tools like GitHub Copilot, offers significant benefits in fostering creativity, enhancing learning efficiency, and supporting advanced projects. In engineering education, GAI offers numerous benefits, leveraging advanced chatbots and text-generation models to enhance learning and problem-solving capabilities. Cloud-based frameworks and social robots significantly enhance engineering education by providing scalable resources, interactive learning environments, and personalized support. Moreover, GAI has the potential to revolutionize medical education by enhancing clinical training, improving diagnostic accuracy, supporting personalized medicine, and advancing public health education. Also, GAI models hold great potential to enhance communication education across journalism, media, and healthcare fields. By supporting content generation, data analysis, creative development, and patient communication, GAI tools can provide valuable learning experiences and improve productivity (Bahroun et al, 2023).

GAI holds immense potential to transform education by enhancing teaching, learning, and educational processes. However, to fully realize these benefits, it is essential to address issues of responsible and ethical usage, potential biases, and academic integrity. By developing comprehensive guidelines, promoting transparency, mitigating bias, and fostering critical thinking skills, educators and institutions can ensure that AI technologies contribute positively to a technologically advanced, inclusive, and effective educational landscape (Bahroun et al, 2023).

*Fostering Intrinsic Motivation through Self-Determination Theory*

SDT posits that students are most motivated when their needs for autonomy, competence, and relatedness are met. GAI can significantly enhance these aspects, fostering intrinsic motivation among students. When students are intrinsically motivated, they are more likely to engage deeply with the material and maintain academic integrity.

AI tools enhance autonomy by allowing students to control their learning process. In a history class, for instance, students can use AI-generated interactive timelines and simulations to explore different aspects of the Industrial Revolution at their own pace. This self-directed exploration encourages students to take ownership of their learning journey, which promotes a genuine interest in the subject matter. Such autonomy reduces the likelihood of dishonest behavior, as students are motivated by curiosity and a desire to learn.

Moreover, AI tools support competence by providing personalized feedback that helps students improve their skills. In a language arts classroom, an AI-driven writing assistant can analyze a student's work and offer specific suggestions for improvement. This real-time feedback not only enhances the student's writing skills but also builds their confidence. When students see tangible improvements in their abilities, their intrinsic motivation to engage with the subject matter increases. This motivation fosters academic integrity, as students take pride in their work and are less inclined to plagiarize or cheat.

GAI also facilitates relatedness by enabling collaborative learning. In project-based learning environments, AI tools can help students work together more effectively. For example, in a science class, an AI-powered lab assistant can guide groups through virtual experiments, encouraging discussion and collaboration. This collaborative process fosters a sense of community and shared purpose among students, which supports their intrinsic motivation to learn and succeed together. When students feel connected to their peers and their learning objectives, they are more likely to adhere to ethical standards and maintain academic integrity.

*Enhancing Digital Literacy and Ethical Awareness*

Digital literacy is essential in today's technology-driven world, and GAI can play a crucial role in fostering this skill. Ethical frameworks such as deontological ethics and consequentialism provide valuable insights into the responsible use of AI in education, emphasizing the importance of honesty, fairness, and positive outcomes.

Deontological ethics, which focuses on adherence to moral principles, underscores the need for using AI tools responsibly. Educators can teach students to use AI ethically by emphasizing principles such as honesty and integrity. For instance, when using AI-generated simulations in a history class, educators can guide students to engage thoughtfully with the material, ensuring that their use of AI supports genuine learning rather than shortcuts. By instilling these ethical values, educators help students understand the importance of maintaining academic integrity.

Consequentialism, which evaluates the morality of actions based on their outcomes, further supports the responsible use of AI in education. The ethical use of AI should aim to produce positive educational outcomes, such as enhanced learning, critical thinking, and digital literacy. In a language arts classroom, an AI writing assistant can provide constructive feedback that helps students refine their writing skills. This positive outcome not only improves their competence but also instills a sense of responsibility in using AI tools ethically. When students see the benefits of using AI to enhance their skills, they are more likely to use these tools responsibly, maintaining academic integrity.

Moreover, educating students on the ethical use of AI tools is crucial for fostering digital literacy. In a science class, an AI-powered lab assistant can guide students through virtual experiments, prompting discussions on ethical considerations such as data privacy and accuracy. By engaging in these discussions, students develop a nuanced understanding of the role of AI in scientific research and the ethical responsibilities that come with it. This awareness empowers students to navigate the digital world ethically and effectively, reducing the likelihood of dishonest behavior.

*Supporting Constructivist Learning Principles*

Constructivist learning theory emphasizes that students construct knowledge through experiences and reflections. GAI aligns well with this theory, offering tools that promote exploration, interaction, and personalized learning paths. By supporting constructivist principles, AI enhances academic integrity by encouraging deeper understanding and critical thinking.

In a history class studying the Industrial Revolution, an AI tool that generates interactive timelines and simulations allows students to manipulate variables and observe outcomes. This hands-on exploration helps students construct a deeper understanding of historical complexities. Rather than passively receiving information, students actively engage with the content, reflecting on the consequences of different actions. This active engagement fosters a genuine interest in learning, reducing the temptation to cheat.

Similarly, in a language arts classroom, a GAI tool that provides real-time feedback on writing helps students improve their narrative skills. By experimenting with different plot developments and character traits, students engage in a creative process that aligns with constructivist principles. This interactive learning experience encourages students to think critically about their stories, fostering a deeper understanding of language and storytelling. When students are genuinely invested in their learning process, they are less likely to engage in dishonest practices.

Collaborative learning, another key aspect of constructivist theory, is also enhanced by GAI. In project-based learning environments, AI tools can facilitate collaboration by organizing information, suggesting relevant sources, and providing feedback on the clarity of students' work. For example, in a business class, an AI tool can help students develop a business plan by generating market analysis reports and financial projections. This collaborative process encourages students to engage in dialogue, share perspectives, and build knowledge collectively. When students work together to achieve common goals, they are more likely to adhere to ethical standards and maintain academic integrity.

**Conclusion**

GAI, when integrated responsibly in education, does not erode academic integrity. Instead, it fosters intrinsic motivation, enhances digital literacy, and supports constructivist learning principles. By promoting autonomy, competence, and relatedness, AI tools help students develop a genuine interest in their subjects, reducing the likelihood of dishonest behavior. Ethical education and personalized feedback further empower students to navigate the digital world responsibly, ensuring that they use AI tools to enhance their learning rather than as shortcuts. Through interactive and collaborative learning experiences, GAI encourages deeper understanding and critical thinking, ultimately promoting academic integrity in today's educational landscape.

*Some Practical Guidance for Educators and Administrators*

To provide practical guidance for using AI in education, we recommend focusing on integrating AI in ways that support established educational goals while adhering to ethical guidelines. Transparency is crucial in this process, as educators must actively involve students in understanding how AI tools are being used, what their limitations are, and why ethical use is important. This includes making the need to understand and actively utilize AI an explicit part of program objectives, course objectives, and learning outcomes, ensuring that its integration aligns with educational goals like developing digital literacy and critical thinking skills. By discussing potential biases, data privacy concerns, and limitations of AI-generated content, educators foster a culture of critical engagement where students learn to use AI responsibly and ethically rather than blindly relying on it. This proactive approach equips students with the discernment and integrity needed to navigate an AI-driven world.

Professional development for educators is crucial for the effective integration of AI in education. Governments and administrative bodies must exert the necessarily sustained and concerted pressures to make this a priority. Alongside sustained and concerted pressures, they need to sufficiently invest in resources and provide support and encouragement to ensure that this training is effective and widespread. Training programs should equip educators with practical skills for using AI tools, while also covering ethical considerations like data privacy, algorithmic bias, and the limitations of AI-generated feedback. By mandating and funding professional development, policymakers and administrators can ensure that educators are well-prepared to navigate the potential risks and benefits of AI. This comprehensive support empowers educators to guide students in using AI tools responsibly, fostering genuine learning and upholding academic integrity, rather than allowing misuse or over-reliance on technology to take root.

Finally, an iterative approach to integrating AI is crucial, and this must be encouraged at the policymaking level as well. Educators should continuously assess the impacts of AI on learning outcomes and be prepared to adjust their strategies accordingly. This involves collecting feedback from students, reviewing the effectiveness of AI tools, and making necessary changes to ensure AI contributes to meaningful educational experiences. Policymakers can support this process by implementing guidelines and providing resources that promote regular evaluation and adaptation of AI integration practices in schools. By emphasizing these practical steps at both the classroom and policy levels, educators can incorporate AI in ways that not only enhance learning but also foster responsible, ethical engagement with technology.